\documentclass[journal,draftclsnofoot,onecolumn,12pt,twoside]{IEEEtranTCOM}
\usepackage[left=1in,top=1in, bottom=1in, right=1in]{geometry}
\date{}
\usepackage{amsmath,cite,amsfonts,amssymb,psfrag,amsthm,paralist}
\usepackage{graphicx}
\usepackage{epstopdf}
\usepackage{rotating}
\usepackage{dsfont}
\usepackage{color}
\usepackage{tikz}
\usetikzlibrary{plotmarks}
\usepackage{pgfplots}
\usetikzlibrary{calc}
\usetikzlibrary{shapes,arrows}
\usetikzlibrary{decorations.markings}
\usetikzlibrary{positioning}
\pgfplotsset{compat=1.10}
\usetikzlibrary{calc}
\usetikzlibrary{shapes,arrows}
\usetikzlibrary{decorations.markings}

\usepackage[utf8]{inputenc}
\usepackage{authblk}

\usepackage{balance}

\DeclareFontFamily{U}{mathx}{\hyphenchar\font45}
\DeclareFontShape{U}{mathx}{m}{n}{<-> mathx10}{}
\DeclareSymbolFont{mathx}{U}{mathx}{m}{n}
\DeclareMathAccent{\widebar}{0}{mathx}{"73}

\def\BibTeX{{\rm B\kern-.05em{\sc i\kern-.025em b}\kern-.08em T\kern-.1667em\lower.7ex\hbox{E}\kern-.125emX}}

\DeclareFontFamily{U}{mathx}{\hyphenchar\font45}
\DeclareFontShape{U}{mathx}{m}{n}{<-> mathx10}{}
\DeclareSymbolFont{mathx}{U}{mathx}{m}{n}

\newcommand{\Enc}{\mathsf{Enc}}
\newcommand{\Dec}{\mathsf{Dec}}
\newcommand{\Est}{\mathsf{Est}}

\makeatletter
\newtheorem*{rep@theorem}{\rep@title}
\newcommand{\newreptheorem}[2]{%
	\newenvironment{rep#1}[1]{%
		\def\rep@title{\Cref{##1}}%
		\begin{rep@theorem}}%
		{\end{rep@theorem}}}
\newcommand*{\textlabel}[2]{%
	\edef\@currentlabel{#1}
	\phantomsection
	#1\label{#2}
}
\makeatother

\usepackage{scalerel}
\usepackage{stackengine}
\stackMath
\def\hatgap{2pt}
\def\subdown{-2pt}
\newcommand\reallywidehat[2][]{%
	\renewcommand\stackalignment{l}%
	\stackon[\hatgap]{#2}{%
		\stretchto{%
			\scalerel*[\widthof{$#2$}]{\kern-.6pt\bigwedge\kern-.6pt}%
			{\rule[-5\textheight]{0.1ex}{\textheight}}
		}{0.5ex}
		_{\smash{\belowbaseline[\subdown]{\scriptstyle#1}}}%
}}

\newtheorem{theorem}{Theorem}

\newtheorem{remark}{Remark}
\newtheorem{definition}{Definition}
\newtheorem{proposition}{Proposition}
\newtheorem{lemma}{Lemma}

\begin{document}
	\title{Secure Joint Communication and Sensing}

\author{Onur G\"unl\"u,~\textit{Member},~\textit{IEEE}, Matthieu Bloch,~\textit{Senior Member},~\textit{IEEE},
	Rafael~F.~Schaefer,~\textit{Senior Member},~\textit{IEEE}, and Aylin Yener,~\textit{Fellow},~\textit{IEEE}
	\thanks{This work has been supported by the German Federal Ministry of Education and Research (BMBF) under the Grant 16KIS1242, German Research Foundation (DFG) under the Grant SCHA 1944/9-1, and National Science Foundation (NSF) under the Grant CCF 1955401.}
	\thanks{O. G\"unl\"u and R. F. Schaefer are with the Chair of Communications Engineering and Security, University of Siegen, 57076 Siegen, Germany (E-mail: \{onur.guenlue, rafael.schaefer\}@uni-siegen.de).}
	\thanks{M. Bloch is with the School of Electrical and Computer Engineering, Georgia Institute of Technology, Atlanta, GA 30332 (E-mail: matthieu.bloch@ece.gatech.edu).}
	\thanks{A. Yener is with the Department of Electrical and Computer Engineering, The Ohio State University, Columbus, OH 43210 (E-mail: yener@ece.osu.edu).}
}

\maketitle

\begin{abstract}
	This work considers the problem of mitigating information leakage between  communication and sensing in systems jointly performing both operations. Specifically, a discrete memoryless state-dependent broadcast channel model is studied in which
	\begin{inparaenum}[(i)]
		\item the presence of feedback enables a transmitter to convey information, while simultaneously performing channel state estimation;
		\item one of the receivers is treated as an eavesdropper whose state should be estimated but which should remain oblivious to part of the transmitted information.
	\end{inparaenum}
	The model abstracts the challenges behind security for joint communication and sensing if one views the channel state as a sensitive attribute, e.g., location. For independent and identically distributed states, perfect output feedback, and when part of the transmitted message should be kept secret, a partial characterization of the secrecy-distortion region is developed. The characterization is exact when the broadcast channel is either physically-degraded or reversely-physically-degraded. The partial characterization is also extended to the situation in which the entire transmitted message should be kept secret. The benefits of a joint approach compared to separation-based secure communication and state-sensing methods are illustrated with a binary joint communication and sensing model.
\end{abstract}

%

\section{Introduction} \label{sec:intro}
The vision for next generation mobile communication networks includes a seamless integration of the physical and digital world. Key to its success is the network's ability to automatically react to changing environments thanks to tight harmonization of communication and sensing~\cite{NokiaGuysJCASTutorial}. For instance, a millimeter wave (mmWave) joint communication and radar system can be used to detect a target or to estimate crucial parameters relevant to communication and adapt the communication scheme accordingly~\cite{JCASwithSecurityTutorial}. Joint communication and sensing (JCAS), or integrated sensing and communication, techniques are envisioned more broadly as key enablers for a wide range of applications, including connected vehicles and drones.

Several information-theoretic studies of JCAS have been initiated, drawing on existing results for joint communication and state estimation~\cite{Zhang_2011,AcademicsJCASTutorial,MassiveMIMOforJCAS, VDEJCASPositionPaper}. Motivated by the integration of communication and radar for mmWave vehicular applications,~\cite{MariGiuseppeGerhardJCAS} considers a model in which messages are encoded and sent through a state-dependent channel with generalized feedback both to reliably communicate with a receiver and to estimate the channel state by using the feedback and transmitted codewords. The optimal trade-off between the communication rate and channel-state estimation distortion is then characterized for memoryless JCAS channels and independent and identically distributed (i.i.d.) channel states that are causally available at the receiver and estimated at the transmitter by using a strictly causal channel output. Follow up works have extended the model to multiple access channels~\cite{MariMACJCAS} and broadcast channels~\cite{MariMicheleBCJCAS}.

The nature of JCAS mandates the use of a single modality for the communication and sensing functions so that sensing signals carry information, which then creates situations in which leakage of sensitive information can occur. For example, a target illuminated for sensing its range has the ability to gather potentially sensitive information about the transmitted message~\cite{SecureJCASWireless}. As the sensing performance and secrecy performance are both measured with respect to the signal received at the sensed target, there exists a trade-off between the two~\cite{JCASwithSecurityTutorial}. To capture and characterize this trade-off, we extend the JCAS model in \cite{MariGiuseppeGerhardJCAS} by introducing an eavesdropper in the network. The objective of the transmitter is then to simultaneously communicate reliably with the legitimate receiver, estimate the channel state, and hide a part of the message from the eavesdropper. The channel state is modeled as a two-component state capturing the characteristics of each individual receiver, the feedback is modeled as perfect output feedback for simplicity, and the transmitted message is divided into two parts, only one of which should be kept (strongly) secret (this is called partial secrecy in~\cite{RaviZivPartialSecrecyWTC}). 

We develop inner and outer bounds on the secrecy-distortion region of this partial-secrecy scenario under a strong secrecy constraint when i.i.d. channel states are causally available at the corresponding receivers. The bounds match when the JCAS channel is physically- or reversely-physically-degraded, and the outer bound also applies to the case of noisy generalized feedback. We also extend these characterizations to the case in which the entire transmitted message should be kept secret. The proposed secure JCAS models can be viewed as extensions of the wiretap channel with feedback models~\cite{AhlswedeCaiWTCwithFeedback,AsafCohenWTCwithFeedback,OurJSAITTutorial,HanVinckWTCwithFeedback,he-yener-fbsecrecy,GermanWTCwithGeneralizedFeedback,YHKimWTCwithFeedback,AminGerhardTwoWaySecrecy}. Our achievability proof leverages the output statistics of random binning (OSRB) method \cite{AhlswedeCsiz,OSRBAmin,RenesRenner} to obtain strong secrecy. A binary JCAS channel example with multiplicative Bernoulli states illustrates how secure JCAS methods may outperform separation-based secure communication and state-sensing methods.

\section{Problem Definition}\label{sec:problem_setting}
\begin{figure}
	\centering
	\resizebox{0.76\linewidth}{!}{
		\begin{tikzpicture}
		\node (a) at (0,-1.0) [draw,rounded corners = 6pt, minimum width=2.2cm,minimum height=0.8cm, align=left] {$\widehat{S^n_j} = \Est_j(X^n,Z^n)$};
		\node (c) at (3.5,-3.1) [draw,rounded corners = 5pt, minimum width=1.3cm,minimum height=0.6cm, align=left] {$P_{Y_1Y_2Z|S_1S_2X}$};
		\node (state) at (7.5,-3.1) [draw,rounded corners = 5pt, minimum width=1.3cm,minimum height=0.6cm, align=left] {$P_{S_1S_2}$};
		\draw[decoration={markings,mark=at position 1 with {\arrow[scale=1.5]{latex}}},
		postaction={decorate}, thick, shorten >=1.4pt] ($(state.west)+(0,0.15)$) -- ($(c.east)+(0,0.15)$) node [near end, above] {$S_{1,i}$};
		\draw[decoration={markings,mark=at position 1 with {\arrow[scale=1.5]{latex}}},
		postaction={decorate}, thick, shorten >=1.4pt] ($(state.west)+(0,-0.15)$) -- ($(c.east)+(0,-0.15)$) node [near end, below] {$S_{2,i}$};
		\node (b) at (4.2,-0.5) [draw,rounded corners = 6pt, minimum width=2.2cm,minimum height=0.8cm, align=left] {$\widehat{M}=\Dec(Y_1^n,S_1^n)$};
		\node (g) at (3.5,-5) [draw,rounded corners = 5pt, minimum width=1cm,minimum height=0.6cm, align=left] {Eve};
		\draw[decoration={markings,mark=at position 1 with {\arrow[scale=1.5]{latex}}},
		postaction={decorate}, thick, shorten >=1.4pt] ($(state.west)+(-1.2,0.15)$) -- ($(b.south)+(1.4,0)$) node [near end, left] {$S_{1,i}$};
		\draw[decoration={markings,mark=at position 1 with {\arrow[scale=1.5]{latex}}},
		postaction={decorate}, thick, shorten >=1.4pt] ($(state.west)+(-1.2,-0.15)$) -- ($(b.south)+(1.47,-4.1)$) --   ($(g.east)+(0,0.00)$) node [near end, above] {$S_{2,i}$};
		\node (a1) [below of = a, node distance = 2.1cm] {$X_i$};
		\draw[decoration={markings,mark=at position 1 with {\arrow[scale=1.5]{latex}}},
		postaction={decorate}, thick, shorten >=1.4pt] ($(c.north)+(0.0,0)$) -- ($(b.south)-(0.7,0)$) node [midway, right] {$Y_{1,i}$};
		\draw[decoration={markings,mark=at position 1 with {\arrow[scale=1.5]{latex}}},
		postaction={decorate}, thick, shorten >=1.4pt] (a1.east) -- ($(c.west)-(0,0.0)$);
		\draw[decoration={markings,mark=at position 1 with {\arrow[scale=1.5]{latex}}},
		postaction={decorate}, thick, shorten >=1.4pt,dashed] (a1.north) -- ($(a.south)-(0,0.0)$);
		\draw[decoration={markings,mark=at position 1 with {\arrow[scale=1.5]{latex}}},
		postaction={decorate}, thick, shorten >=1.4pt] ($(c.north)-(0.5,0.0)$) -- ($(c.north)-(0.5,-0.3)$) -- ($(c.north)-(4.5,-0.3)$) -- ($(a.south)-(1,0)$);
		\draw[decoration={markings,mark=at position 1 with {\arrow[scale=1.5]{latex}}},
		postaction={decorate}, thick, shorten >=1.4pt] (c.south) -- (g.north) node [midway, right] {$Y_{2,i}$};
		\node (b2) [right of = b, node distance = 4cm] {$\widehat{M}=\big(\widehat{M}_1,\widehat{M}_2\big)$};
		\draw[decoration={markings,mark=at position 1 with {\arrow[scale=1.5]{latex}}},
		postaction={decorate}, thick, shorten >=1.4pt] (b.east) -- (b2.west);
		\node (a2) [below of = a, node distance = 5cm] {$M=(M_1,M_2)$};
		\node (f2) at (0,-4.5) [draw,rounded corners = 5pt, minimum width=1cm,minimum height=0.6cm, align=left] {$X_i=\Enc_i(M,Z^{i-1})$};
		\draw[decoration={markings,mark=at position 1 with {\arrow[scale=1.5]{latex}}},
		postaction={decorate}, thick, shorten >=1.4pt]  (f2.north) -- (a1.south);
		\draw[decoration={markings,mark=at position 1 with {\arrow[scale=1.5]{latex}}},
		postaction={decorate}, thick, shorten >=1.4pt] (a2.north) -- (f2.south) ;
		\draw[decoration={markings,mark=at position 1 with {\arrow[scale=1.5]{latex}}},
		postaction={decorate}, thick, shorten >=1.4pt] ($(c.north)-(4.5,-0.3)$) -- ($(f2.north)-(1,0)$) node [pos=0.0, left] {$Z_{i-1}$};
		\end{tikzpicture}
	}
	\caption{JCAS model with partial secrecy, where only $M_2$ should be kept secret from Eve, for $j=1,2$ and $i~=~[1:n]$. We mainly consider JCAS with perfect output feedback, where $Z_{i-1}=(Y_{1,i-1},Y_{2,i-1})$.}\label{fig:SecureJCASModel}
\end{figure}
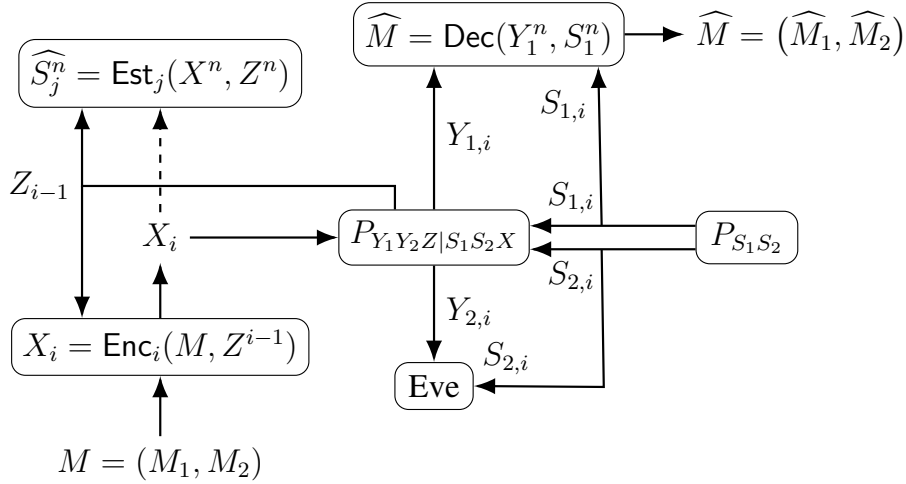

We consider the secure JCAS model shown in Fig.~\ref{fig:SecureJCASModel}, which includes a transmitter equipped with a state estimator, a legitimate receiver, and an eavesdropper (Eve). The transmitter attempts to reliably transmit a uniformly distributed message $M=(M_1,M_2)\in \mathcal{M}=\mathcal{M}_1\times \mathcal{M}_2$ through a memoryless state-dependent JCAS channel with known statistics $P_{Y_1Y_2Z|S_1S_2X}$ and i.i.d. state sequence $(S_1^n,S_2^n)\in\mathcal{S}_1^n\times\mathcal{S}_2^n$ generated according to a known joint probability distribution $P_{S_1S_2}$. The transmitter calculates the channel inputs $X^n$ as $X_i=\Enc_i(M,Z^{i-1})\in \mathcal{X}$ for all $i=[1:n]$, where $\Enc_i(\cdot)$ is an encoding function and $Z^{i-1}\in\mathcal{Z}^{i-1}$ is the delayed channel output feedback. The legitimate receiver that observes $Y_{1,i}\in\mathcal{Y}_1$ and $S_{1,i}$ for all channel uses $i=[1:n]$ should reliably decode both $M_1$ and $M_2$ by forming the estimate $\widehat{M}=\Dec(Y_1^n,S_1^n)$, where $\Dec(\cdot)$ is a decoding function. The eavesdropper that observes $Y_{2,i}\in\mathcal{Y}_2$ and $S_{2,i}$ should be kept ignorant of $M_2$. Finally, the transmitter estimates the state sequence $(S_1^n,S_2^n)$ as $\widehat{S^n_j} = \Est_j(X^n,Z^n)\in\reallywidehat[n]{\mathcal{S}_j}$ for $j=1,2$, where $\Est_j(\cdot,\cdot)$ is an estimation function. Unless specified otherwise, all sets $\mathcal{S}_1$, $\mathcal{S}_2$, $\widehat{\mathcal{S}}_1$, $\widehat{\mathcal{S}}_2$, $\mathcal{X}$, $\mathcal{Y}_1$, $\mathcal{Y}_2$, and $\mathcal{Z}$ are finite.

For simplicity, we consider the perfect output feedback case in which for all $i=[2:n]$ we have
\begin{align}
	Z_{i-1}=(Y_{1,i-1},Y_{2,i-1}).\label{eq:POFcondition}
\end{align}
Although this is explicitly used in our achievability proofs, some of our converse results hold for generalized feedback. We next define the strong secrecy-distortion region for the problem of interest.

\begin{definition}\label{def:systemmodel}
	\normalfont A secrecy-distortion tuple $(R_{1}, R_{2},D_{1},D_{2})$ is \emph{achievable} if, for any $\delta\!>\!0$, there exist $n\!\geq\!1$, one encoder, one decoder, and two estimators $\Est_j(X^n,Y_1^n,Y_2^n )=\widehat{S_j^n}$ for $j=1,2$ such that
	\begin{align}
		& \frac{1}{n}\log |\mathcal{M}_j|\geq R_j -\delta\quad\;\;\;\;\;\text{for } j\!=\!1,2\;\;&&\!\!\!\!\!(\text{rates})\label{eq:rates_cons}\\ 
		&\Pr\big[M \neq \widehat{M}\big] \leq \delta&&\!\!\!\!\! (\text{reliability})\label{eq:reliability_cons}\\
		&I(M_2;Y^n_2|S_2^n) \leq \delta&&\!\!\!\!\!(\text{strong secrecy})\label{eq:secrecyleakage_cons}\\
		&\mathbb{E}\big[d_j(S_j^n,\widehat{S_j^n})\big] \!\leq\! D_j\!+\!\delta\;\;\;\;\;\;\text{for } j\!=\!1,2\;\;&&\!\!\!\!\!(\text{distortions})\label{eq:distortion_consts}
	\end{align}
	where $d_j(s^n,\widehat{s^n})=\frac{1}{n}\sum_{i=1}^nd_j(s_i,\widehat{s}_i)$ for $j\!=\!1,2$ are bounded per-letter distortion metrics.
	
	The secrecy-distortion region $\mathcal{R}_{\textnormal{PS,POF}}$ is the closure of the set of all achievable tuples with partial secrecy and perfect output feedback. \hfill $\lozenge$
\end{definition}

The use of per-letter distortion metrics $d_j(\cdot,\cdot)$ in conjunction with i.i.d. states simplifies the problem to a rate distortion region characterization \cite{MariGiuseppeGerhardJCAS,MariMicheleBCJCAS,MariMACJCAS}; in fact, past observations are independent of present and future ones, lending the transmitter no state prediction ability to adapt its transmission on the fly. Analyzing JCAS models with memory leads to conceptually different results; see, e.g.,~\cite{Chang2022Rate}.

\begin{remark}\label{rem:secrecyconstraintwithoutcond}
	\normalfont The strong secrecy condition (\ref{eq:secrecyleakage_cons}) is equivalent to $I(M_2;Y^n_2,S_2^n) \leq \delta$ since the transmitted message is independent of the state sequence.
\end{remark}

\section{Bounds for JCAS with Partial-Secrecy}\label{sec:JCASwithPSandPOFResults}
We next provide inner and outer bounds on the secrecy-distortion region $\mathcal{R}_{\textnormal{PS,POF}}$; see Section~\ref{sec:proofinnerouterPSPOF} for a proof sketch.

Define $[a]^+=\max\{a,0\}$ for $a\in\mathbb{R}$.

\begin{proposition}[Inner Bound]\label{prop:InnerforPSPOF}
	The region $\mathcal{R}_{\textnormal{PS,POF}}$ includes the union over all joint distributions $P_{UVX}$ of the rate tuples $(R_{1}, R_{2},D_1,D_2)$ such that
	\begin{align}
		&R_{1}\leq I(U;Y_1|S_1)\label{eq:achR1}\\
		&R_{2}\leq \min\{R_{2}^{\prime},\quad (I(V;Y_1|S_1)-R_1)\}\label{eq:achR2}\\
		& D_j\geq \mathbb{E}[d_j(S_j,\widehat{S}_j))]\qquad\qquad  \text{for }j=1,2\label{eq:achdistortion1and2}
	\end{align}
	where
	\begin{align}
		&P_{UVXY_1Y_2S_1S_2} = P_{U|V}P_{V|X}P_XP_{S_1S_2}P_{Y_1Y_2|S_1S_2X}\label{eq:jointprobiid},\\
		&R_{2}^{\prime}=[I(V;Y_1|S_1,U)-I(V;Y_2|S_2,U)]^{+}+H(Y_1|Y_2,S_2,V)\label{eq:R2primedef}
	\end{align}
	and one can apply the per-letter estimators $\Est_j(x,y_1,y_2)=\hat{s}_j$ for $j=1,2$ such that 
	\begin{align}
		\Est_j(x,y_1,y_2)=\mathop{\textnormal{argmin}}_{\tilde{s}\in\widehat{\mathcal{S}}_j} \sum_{s_j\in\mathcal{S}_j}P_{S_j|XY_1Y_2}(s_j|x,y_1,y_2)\; d_j(s_j,\tilde{s}).\label{eq:deterministicest} 
	\end{align}
	One can limit $|\mathcal{U}|$ to 
	\begin{align}
		\min\{|\mathcal{X}|,\;|\mathcal{Y}_1|\!\cdot\!|\mathcal{S}_1|,\;|\mathcal{Y}_2|\!\cdot\!|\mathcal{S}_2|\}\!+\!2\label{eq:cardUforPS}
	\end{align}
	and $|\mathcal{V}|$ to
	\begin{align}
		(\min\{|\mathcal{X}|,\;|\mathcal{Y}_1|\!\cdot\!|\mathcal{S}_1|,\;|\mathcal{Y}_2|\!\cdot\!|\mathcal{S}_2|\}\!+\!2)\cdot(\min\{|\mathcal{X}|,\;|\mathcal{Y}_1|\!\cdot\!|\mathcal{S}_1|,\;|\mathcal{Y}_2|\!\cdot\!|\mathcal{S}_2|\}\!+\!1).\label{eq:cardVforPS}
	\end{align}
\end{proposition}

\begin{proposition}[Outer Bound]\label{prop:OuterforPSPOF}
	The region $\mathcal{R}_{\textnormal{PS,POF}}$ is included in the union over all joint distributions $P_{UVX}$ of the rate tuples $(R_1,R_2,D_1,D_2)$ satisfying (\ref{eq:achdistortion1and2}) and
	\begin{align}
		&R_{1}\leq I(V;Y_1|S_1)\label{eq:achR1degradedbutnowouterProp2}\\
		&R_2\leq \min\Big\{\big(H(Y_1,S_1|Y_2,S_2) \!-\! H(S_1|Y_1,Y_2,S_2,V)\big),\quad \big(I(V;Y_1|S_1)-R_1\big)\Big\}    \label{eq:convR2}
	\end{align}
	where we have (\ref{eq:jointprobiid}) and (\ref{eq:deterministicest}). One can limit $|\mathcal{V}|$ to
	\begin{align}
				\min\{|\mathcal{X}|,\;|\mathcal{Y}_1|\!\cdot\!|\mathcal{S}_1|,\;|\mathcal{Y}_2|\!\cdot\!|\mathcal{S}_2|\}\!+\!1.\label{eq:Vcard+1}
	\end{align}
		
\end{proposition}

\begin{remark}
	\normalfont Since  we consider perfect feedback as in~(\ref{eq:POFcondition}), the outer bound proposed in Proposition~\ref{prop:OuterforPSPOF} is also valid for the general JCAS problem depicted in Fig.~\ref{fig:SecureJCASModel}, in which the feedback $Z_{i-1}$ can be a noisy version of $(Y_{1,i-1},Y_{2,i-1})$.
\end{remark}

We next characterize the strong secrecy-distortion regions of physically-degraded and reversely-physically-degraded JCAS channels with partial secrecy and perfect output feedback, defined below; see also \cite[Definition~2]{MariMicheleBCJCAS}. 

\begin{definition}\label{def:physicallydegraded}
	\normalfont A JCAS channel $P_{Y_1Y_2|S_1S_2X}$ is \emph{physically-degraded} if we have
	\begin{align}
		P_{Y_1Y_2S_1S_2|X}=P_{Y_1Y_2|S_1S_2X}P_{S_1S_2}=P_{S_1}P_{Y_1|S_1X}P_{Y_2S_2|S_1Y_1}\label{eq:physicaldegradedcond}
	\end{align}
	and is \emph{reversely-physically-degraded} if the degradation order is changed such that
	\begin{align}
		P_{Y_1Y_2S_1S_2|X}=P_{Y_1Y_2|S_1S_2X}P_{S_1S_2}=P_{S_2}P_{Y_2|S_2X}P_{Y_1S_1|S_2Y_2}.\label{eq:reverselyphysicaldegradedcond}
	\end{align}\hfill $\lozenge$
\end{definition}
A physically-degraded JCAS channel corresponds to a situation in which the observations $(Y_2^n,S_2^n)$ of the eavesdropper are degraded versions of observations $(Y_1^n,S_1^n)$ of the legitimate receiver with respect to the channel input $X^n$. 

\begin{theorem}\label{theo:SimplifiedforPSPOFDegraded}
	{\normalfont(Physically-degraded Channels):} For a physically-degraded JCAS channel, $\mathcal{R}_{\textnormal{PS,POF}}$ is the union over all joint distributions $P_{VX}$ of the rate tuples $(R_{1}, R_{2},D_1,D_2)$ satisfying (\ref{eq:achdistortion1and2}), (\ref{eq:achR1degradedbutnowouterProp2}), and (\ref{eq:convR2}), where we have (\ref{eq:jointprobiid}) with constant $U$ and (\ref{eq:deterministicest}). One can limit $|\mathcal{V}|$ to (\ref{eq:Vcard+1}).
\end{theorem}

\begin{IEEEproof}[Proof of Theorem~\ref{theo:SimplifiedforPSPOFDegraded}]
	Since the outer bound given in Proposition~\ref{prop:OuterforPSPOF} does not assume any degradedness, the outer bound terms for $R_1$, $R_2$, and $D_j$ for $j=1,2$ follow from Proposition~\ref{prop:OuterforPSPOF}.

	The achievability proof for Theorem~\ref{theo:SimplifiedforPSPOFDegraded} follows by modifying the proof of Proposition~\ref{prop:InnerforPSPOF}. We next provide a sketch of the modifications for a physically-degraded JCAS channel. First, $U^n$ is not used, i.e., $U^n$ is eliminated from the achievability proof. Second, to each $v^n(k)$ we assign four random bin indices $(F_{\text{v}}(k),W_{\text{v}_1}(k),W_{\text{v}_2}(k),L_{\text{v}}(k))$ such that $F_{\text{v}}(k)\in[1:2^{n\widetilde{R}_{\text{v}}}]$, $W_{\text{v}_1}(k)\in[1:2^{nR_{\text{v}_1}}]$, $W_{\text{v}_2}(k)\in[1:2^{nR_{\text{v}_2}}]$, and $L_{\text{v}}(k)\in[1:2^{n\overline{R}_{\text{v}}}]$ for all $k=[1:b]$ independently such that $M_1(k)=W_{\text{v}_1}(k)$ and $M_2(k)=(W_{\text{v}_2}(k),L_{\text{v}}(k))$. As in (\ref{eq:reconstrV}), we impose the reliability constraint
	\begin{align} 
		\widetilde{R}_{\text{v}} > H(V|Y_1,S_1)\label{eq:theo2reliability}
	\end{align}
	as in (\ref{eq:independenceofFvWv}) and (\ref{eq:independenceofYindexnewwwwwwww}) we impose the strong secrecy constraints
	\begin{align}
		&R_{\text{v}_2}+\widetilde{R}_{\text{v}}<H(V|Y_2,S_2)\\
		&\overline{R}_{\text{v}}<H(Y_1|Y_2,S_2,V)
	\end{align} 
	and as in (\ref{eq:sumindependence}) we impose the mutual independence and uniformity constraint
	\begin{align}
		R_{\text{v}_1}+R_{\text{v}_2}+\widetilde{R}_{\text{v}} + \overline{R}_{\text{v}}< H(V).\label{eq:theo2independence}
	\end{align}
	
	We remark that we have $H(V|Y_2,S_2)\geq H(V|Y_1,S_1)$ for all physically-degraded JCAS channels, i.e., we obtain
	\begin{align}
		&[I(V;Y_1|S_1)-I(V;Y_2|S_2)]^{+}\overset{(a)}{=} H(V|Y_2,S_2)-H(V|Y_1,S_1)\label{eq:plussignfordegraded}
	\end{align}
	where $(a)$ follows because $V$ is independent of $(S_1,S_2)$ and since 
	\begin{align}
		V-X-(Y_1,S_1)-(Y_2,S_2)\label{eq:MarkovVXY1S1Y2S2}
	\end{align}
	form a Markov chain for such JCAS channels. Define
	\begin{align}
		&R_{2,\text{deg}}^{\prime} =[I(V;Y_1|S_1)-I(V;Y_2|S_2)]^{+}+H(Y_1|Y_2,S_2,V)\nonumber\\
		&\overset{(a)}{=} H(V|Y_2,S_2)-H(V|Y_1,S_1)+H(Y_1|Y_2,S_2,V)\nonumber\\
		&\overset{(b)}{=}H(Y_1,V|Y_2,S_2)-H(V|Y_1,S_1,Y_2,S_2)\nonumber\\
		&= H(Y_1|Y_2,S_2)+I(V;S_1|Y_1,Y_2,S_2)\nonumber\\
		&= H(Y_1,S_1|Y_2,S_2)- H(S_1|Y_1,Y_2,S_2,V)
	\end{align}
	where $(a)$ follows by (\ref{eq:plussignfordegraded}) and $(b)$ follows from the Markov chain in (\ref{eq:MarkovVXY1S1Y2S2}). 
	
	Applying the Fourier-Motzkin elimination \cite{FMEbook} to (\ref{eq:theo2reliability})-(\ref{eq:theo2independence}), for any $\epsilon>0$ one can achieve 
	\begin{align}
		&R_1= R_{\text{v}_1} =I(V;Y_1,S_1)-2\epsilon=I(V;Y_1|S_1)-2\epsilon\label{eq:R1forTheo2}
	\end{align}
	and for any $R_1$ that is less than or equal to (\ref{eq:R1forTheo2}), one can simultaneously achieve
	\begin{align}
		&R_2 = R_{\text{v}_2} +\overline{R}_{\text{v}} = \min\{R_{2,\text{deg}}^{\prime} ,\quad (I(V;Y_1|S_1)-R_1)\}-3\epsilon.
	\end{align}
	Furthermore, the proofs for achievable distortions, sufficiency of given deterministic estimators, inversion of the problem in the source model into the problem in the  channel model, and elimination of the public indices follow similarly as in the proof of Proposition~\ref{prop:InnerforPSPOF}, so we omit them.
\end{IEEEproof}

\begin{theorem}\label{theo:reversePSPOF}
	{\normalfont(Reversely-physically-degraded Channels):} For a reversely-physically-degraded JCAS channel, $\mathcal{R}_{\textnormal{PS,POF}}$ is the union over all joint distributions $P_{VX}$ of the rate tuples $(R_{1}, R_{2},D_1,D_2)$ satisfying (\ref{eq:achdistortion1and2}), (\ref{eq:achR1degradedbutnowouterProp2}), and 
	\begin{align}
		R_2\leq\min\big\{H(Y_1|Y_2,S_2),\quad \big(I(V;Y_1|S_1)-R_1\big)\big\}   
	\end{align}
	where we have (\ref{eq:jointprobiid}) with constant $U$ and (\ref{eq:deterministicest}). One can limit $|\mathcal{V}|$ to 
	\begin{align}
		\min\{|\mathcal{X}|,\;|\mathcal{Y}_1|\!\cdot\!|\mathcal{S}_1|,\;|\mathcal{Y}_2|\!\cdot\!|\mathcal{S}_2|\}.\label{eq:Vcardinalityforreversephydegraded}
	\end{align}
	
\end{theorem}

\begin{IEEEproof}[Proof of Theorem~\ref{theo:reversePSPOF}]
	The achievability proof follows from Proposition~\ref{prop:InnerforPSPOF} after elimination of $U$ from its proof, as in the proof for Theorem~\ref{theo:SimplifiedforPSPOFDegraded}. After removal of $U$, by (\ref{eq:achR2}) we have the inner bound
	\begin{align}
		&R_2\overset{(a)}{\leq} \min \big\{H(Y_1|Y_2,S_2,V),\quad \big(I(V;Y_1|S_1)-R_1\big)\big\}\nonumber\\
		&\overset{(b)}{=}   \min \big\{H(Y_1|Y_2,S_2),\quad \big(I(V;Y_1|S_1)-R_1\big)\big\}
	\end{align}
	where $(a)$ follows since $V$ is independent of $(S_1,S_2)$ and because $H(V|Y_1,S_1)\geq H(V|Y_2,S_2)$ for all reversely-physically-degraded JCAS channels because of the Markov chain 
	\begin{align}
		V-X-(Y_2,S_2)-(Y_1,S_1)\label{eq:MarkovVXY2S2Y1S1}
	\end{align}
	and $(b)$ follows also because of the Markov chain in (\ref{eq:MarkovVXY2S2Y1S1}).

	Since the outer bound in Proposition~\ref{prop:OuterforPSPOF} does not assume any degradedness, the outer bound terms for $R_1$ and $D_j$ for $j=1,2$ follow from Proposition~\ref{prop:OuterforPSPOF}. Furthermore, by (\ref{eq:convR2}) we obtain the outer bound
	\begin{align}
		&R_2\overset{(a)}{\leq} \min\big\{\big(H(Y_1,S_1|Y_2,S_2) - H(S_1|Y_1,Y_2,S_2)\big),\quad  \big(I(V;Y_1|S_1)-R_1\big)\big\}  \nonumber\\
		&=   \min \big\{H(Y_1|Y_2,S_2),\quad \big(I(V;Y_1|S_1)-R_1\big)\big\}
	\end{align}
	where $(a)$ follows from the Markov chain in (\ref{eq:MarkovVXY2S2Y1S1}).
\end{IEEEproof}

\section{Bounds for JCAS with Single Secure Message}\label{sec:SingleMess}
We next give inner and outer bounds for the situation, in which $M=M_2$ should be kept secret from the eavesdropper and $M_1=\varnothing$. For this situation, the definitions of an achievable secrecy-distortion tuple $(R,D_1,D_2)$ and corresponding strong secrecy-distortion region $\mathcal{R}_{\textnormal{POF}}$ follow from Definition~\ref{def:systemmodel} by eliminating $(M_1,R_1)$ and replacing $(M_2,R_2, \mathcal{R}_{\textnormal{PS,POF}})$ with $(M,R,\mathcal{R}_{\textnormal{POF}})$, respectively.

\begin{proposition}\label{prop:InnerforPOF}
	\emph{(Inner Bound):} The region $\mathcal{R}_{\textnormal{POF}}$ includes the union over all joint distributions $P_{VX}$ of the rate tuples $(R,D_1,D_2)$ satisfying (\ref{eq:achdistortion1and2}) and
	\begin{align}
		&R\leq \min\{R^{\prime\prime},\quad I(V;Y_1|S_1)\}\label{eq:achRforPOF}
	\end{align}
	where 
	\begin{align}
		&P_{VXY_1Y_2S_1S_2} = P_{V|X}P_XP_{S_1S_2}P_{Y_1Y_2|S_1S_2X},\label{eq:jointprobiidforPOF}\\
		&R^{\prime\prime}=[I(V;Y_1|S_1)-I(V;Y_2|S_2)]^{+}+H(Y_1|Y_2,S_2,V)\label{eq:R2doubleprime}
	\end{align}
	and one can apply the deterministic per-letter estimators in (\ref{eq:deterministicest}). One can limit $|\mathcal{V}|$ to (\ref{eq:Vcard+1}).
\end{proposition}

\begin{IEEEproof}[Proof of Proposition~\ref{prop:InnerforPOF}]
	The proof follows by eliminating $U$ in the proof of Proposition~\ref{prop:InnerforPSPOF}, so $R_1=R_{\text{v}_1}=0$ and by imposing (\ref{eq:theo2reliability})-(\ref{eq:theo2independence}) after replacing $R_{\text{v}_2}$ with $R_{\text{v}}$, since for this case we have $M(k)=(W_{\text{v}}(k),L_{\text{v}}(k))$ for all $k=[1:b]$.
\end{IEEEproof}

\begin{proposition}\label{prop:OuterforPOF}
	\emph{(Outer Bound):} The region $\mathcal{R}_{\textnormal{POF}}$ is included in the union over all $P_X$ of the rate tuples $(R,D_1,D_2)$ satisfying (\ref{eq:achdistortion1and2}) and
	\begin{align}
		&R\leq \min\Big\{\big(H(Y_1,S_1|Y_2,S_2) - H(S_1|Y_1,Y_2,S_2,X)\big), \quad I(X;Y_1|S_1)\Big\}    \label{eq:convRforPOF}
	\end{align}
	where one can apply the deterministic per-letter estimators in (\ref{eq:deterministicest}). 
\end{proposition}

\begin{IEEEproof}[Proof of Proposition~\ref{prop:OuterforPOF}]
	Assume that for some $\delta_n\!>\!0$ and $n\geq 1$, there exist an encoder, a decoder, and estimators such that all constraints imposed on the JCAS problem with perfect output feedback are satisfied for some tuple $(R,D_1,D_2)$. We then obtain
	\begin{align}
		&nR\overset{(a)}{\leq} I(M;Y_1^n|S_1^n)+n\epsilon_n\nonumber\\
		&\leq \sum_{i=1}^n \big(H(Y_{1,i}|S_{1,i}) - H(Y_{1,i}|Y_1^{i-1},S_1^n,M,X_i)+\epsilon_n\big)\nonumber\\
		&\overset{(b)}{=} \sum_{i=1}^n \big(H(Y_{1,i}|S_{1,i}) - H(Y_{1,i}|S_{1,i},X_i)+\epsilon_n\big)\nonumber\\
		&= \sum_{i=1}^n (I(X_i;Y_{1,i}|S_{1,i})+\epsilon_n)\label{eq:outerboundforsinglemessageIXY1givenS1}
	\end{align}
	where $(a)$ follows because $M$ and $S_1^n$ are independent, and from Fano's inequality for an $\epsilon_n>0$ such that $\epsilon_n\rightarrow 0$ if $\delta_n\rightarrow 0$, which is entirely similar to (\ref{eq:fanoapp}), and $(b)$ follows because 
	\begin{align}
		Y_{1,i}-(S_{1,i},X_i)-(Y_1^{i-1},S_1^{n\setminus i},M)
	\end{align}
	form a Markov chain. Furthermore, we also have
	\begin{align}
		&nR\overset{(a)}{\leq}I(M;Y_1^n,Y_2^n,S_1^n,S_2^n)+n\epsilon_n\nonumber\\
		&= H(Y_1^n,S_1^n|Y_2^n,S_2^n)+I(Y_2^n,S_2^n;M) -H(Y_1^n,S_1^n|Y_2^n,S_2^n,M)+n\epsilon_n\nonumber\\
		&\overset{(b)}{\leq} \sum_{i=1}^nH(Y_{1,i},S_{1,i}|Y_{2,i},S_{2,i})+ \delta_n -\sum_{i=1}^nH(S_{1,i}|Y_1^n,Y_{2}^n,S_{2}^n,M,S_1^{i-1},X_i)+n\epsilon_n\nonumber\\
		&\overset{(c)}{=}\sum_{i=1}^n\Big(H(Y_{1,i},S_{1,i}|Y_{2,i},S_{2,i})-H(S_{1,i}|Y_{1,i},Y_{2,i},S_{2,i},X_i)+\epsilon_n\Big)+\delta_n
	\end{align}
	where $(a)$ follows from Fano's inequality, which is similar to (\ref{eq:fanoapp}), $(b)$ follows by (\ref{eq:secrecyleakage_cons}) and from Remark~\ref{rem:secrecyconstraintwithoutcond} after replacing $M_2$ with $M$ for the JCAS problem with a single secure message, and $(c)$ follows because 
	\begin{align}
		S_{1,i}-(Y_{1,i},Y_{2,i},S_{2,i},X_i)-(Y_1^{n\setminus i},Y_2^{n\setminus i}, S_2^{n\setminus i},M,S_1^{i-1})\label{eq:Markovsinglesecmessageouternew}
	\end{align}
	form a Markov chain. Thus, by applying the distortion bound in (\ref{eq:outerbounddistortion}) and introducing a uniformly-distributed time-sharing random variable, as being applied in the proof of Proposition~\ref{prop:OuterforPSPOF}, we prove the outer bound for the JCAS problem with a single secure message and perfect output feedback by letting $\delta_n\rightarrow 0$.
\end{IEEEproof} 

We next present the exact strong secrecy-distortion regions for the JCAS problem with a single secure message when the JCAS channel $P_{Y_1Y_2|S_1S_2X}$ is physically-degraded, as in (\ref{eq:physicaldegradedcond}), or reversely-physically-degraded, as in (\ref{eq:reverselyphysicaldegradedcond}).

\begin{theorem}\label{theo:SimplifiedforPOF}
	{\normalfont(Physically-degraded Channels):} For a physically-degraded JCAS channel, $\mathcal{R}_{\textnormal{POF}}$ is the union over all probability distributions $P_{X}$ of the rate tuples $(R,D_1,D_2)$ satisfying (\ref{eq:achdistortion1and2}) and (\ref{eq:convRforPOF}), where we have (\ref{eq:deterministicest}). 
\end{theorem}

\begin{IEEEproof} [Proof of Theorem~\ref{theo:SimplifiedforPOF}]
	Since the bound given in Proposition~\ref{prop:OuterforPOF} is valid for any JCAS channel, the proof for the outer bound follows from Proposition~\ref{prop:OuterforPOF}. Furthermore, the achievability proof follows by modifying the proof of Theorem~\ref{theo:SimplifiedforPSPOFDegraded} such that we assign $V^n(k)=X^n(k)$ for all $k=[1:b]$ and then apply the same OSRB steps for $X^n(k)$ rather than $V^n(k)$, i.e., replace $V$ with $X$ in the inner bound terms given in Proposition~\ref{prop:InnerforPOF}. Define
	\begin{align}
		&R^{\prime\prime}_{\text{deg}}=[I(X;Y_1|S_1)-I(X;Y_2|S_2)]^{+}+H(Y_1|Y_2,S_2,X)\nonumber\\
		&\overset{(a)}{=}I(X;Y_1,S_1|Y_2,S_2)+H(Y_1|Y_2,S_2,X)\nonumber\\
		&=H(Y_1,S_1|Y_2,S_2)-H(S_1|Y_1,Y_2,S_2,X)
	\end{align}
	where $(a)$ follows because the JCAS channel is physically-degraded, and since $X$ is independent of $(S_1,S_2)$. Thus, by (\ref{eq:achRforPOF}) we have
	\begin{align}
		&R\leq \min\{R^{\prime\prime}_{\text{deg}},\quad I(X;Y_1|S_1)\}
	\end{align}
	which proves the achievability bound.
\end{IEEEproof}

\begin{theorem}\label{theo:reversePOFsimplified}
	{\normalfont(Reversely-physically-degraded Channels):} For a reversely-physically-degraded JCAS channel, $\mathcal{R}_{\textnormal{POF}}$ is the union over all probability distributions $P_{X}$ of the rate tuples $(R,D_1,D_2)$ satisfying (\ref{eq:achdistortion1and2}) and
	\begin{align}
		R\leq\min\big\{H(Y_1|Y_2,S_2),\quad I(X;Y_1|S_1)\big\}   
	\end{align}
	where one can apply the deterministic per-letter estimators in (\ref{eq:deterministicest}).	
\end{theorem}

\begin{IEEEproof}[Proof of Theorem~\ref{theo:reversePOFsimplified}]
	We assign $V^n=X^n$ in the achievability proof, i.e., we choose $V=X$ that is allowed by (\ref{eq:jointprobiidforPOF}), such that by (\ref{eq:achRforPOF}) we obtain the inner bound
	\begin{align}
		&R\overset{(a)}{\leq} \min \big\{H(Y_1|Y_2,S_2,X),\quad I(X;Y_1|S_1)\big\}\nonumber\\
		&\overset{(b)}{=}   \min \big\{H(Y_1|Y_2,S_2),\quad I(X;Y_1|S_1)\big\}
	\end{align}
	where $(a)$ follows since $X$ is independent of $(S_1,S_2)$ and because $H(X|Y_1,S_1)\geq H(X|Y_2,S_2)$ for all reversely-physically-degraded JCAS channels due to the Markov chain in (\ref{eq:MarkovVXY2S2Y1S1}), and $(b)$ follows also because of the Markov chain in (\ref{eq:MarkovVXY2S2Y1S1}).

	Since the outer bound in Proposition~\ref{prop:OuterforPOF} does not assume any degradedness, the outer bound terms for $D_j$ for $j=1,2$ follow from Proposition~\ref{prop:OuterforPOF}. Furthermore, by (\ref{eq:convRforPOF}) we have the outer bound
	\begin{align}
		&R\overset{(a)}{\leq} \min\Big\{\big(H(Y_1,S_1|Y_2,S_2) - H(S_1|Y_1,Y_2,S_2)\big),\quad  I(X;Y_1|S_1)\Big\}  \nonumber\\
		&=   \min \big\{H(Y_1|Y_2,S_2),\quad I(X;Y_1|S_1)\big\}
	\end{align}
	where $(a)$ follows from the Markov chain in (\ref{eq:MarkovVXY2S2Y1S1}).
\end{IEEEproof}

\section{Binary JCAS Channel with Multiplicative Bernoulli States Example}\label{sec:JCASexample1}
We next consider a scenario with perfect output feedback and single secure message, in which channel input and output alphabets are binary with multiplicative Bernoulli states, which serves as a coarse model of fading channels with high signal-to-noise ratio. Specifically, we have
\begin{align}
	Y_1 =S_1\cdot X,\qquad\qquad\qquad Y_2=S_2\cdot X
\end{align}
and
\begin{align}
	P_{S_1S_2}(0,0)\!=\!(1\!-\!q), \qquad
	P_{S_1S_2}(1,1)\!=\!q\alpha,\qquad
	P_{S_1S_2}(0,1)\!=\!0,\qquad
	P_{S_1S_2}(1,0)\!=\!q(1\!-\!\alpha)\label{eq:PS1S2exampledistribution} 
\end{align}
for fixed $q,\alpha\in[0,1]$, so the JCAS channel satisfies (\ref{eq:physicaldegradedcond}) \cite[Section~IV-A]{MariMicheleBCJCAS}. 

Define the binary entropy function $H_b(x)=-x\log(x)-(1-x)\log(1-x)$ and denote a Bernoulli random variable $X$ with probability $p$ of success as $X\sim\text{Bern}(p)$.

\begin{lemma}\label{lem:firstexampleregion}
	The strong secrecy-distortion region $\mathcal{R}_{\textnormal{POF}}$ for a binary JCAS channel with multiplicative Bernoulli states characterized by parameters $(q,\alpha)$ and with Hamming distortion metrics is the union over all $p\in[0,1]$, where $X\sim\text{Bern}(p)$, of the rate tuples  $(R,D_1,D_2)$ satisfying
	\begin{align}
		&R\leq \min\Bigg\{\Bigg(q(1-\alpha)H_b(p)+ p(1-q\alpha)H_b\Big(\frac{q(1-\alpha)}{(1-q\alpha)}\Big)\Bigg), \quad qH_b(p)\Bigg\}\label{eq:rateforexample}\\
		&D_1\geq (1-p)\cdot \min\{q,\; (1-q)\}\label{eq:distortion1forexample}\\
		&D_2\geq (1-p)\cdot\min\{q\alpha,\; (1-q\alpha)\}.\label{eq:distortion2forexample}
	\end{align}
\end{lemma}

\begin{IEEEproof}[Proof of Lemma~\ref{lem:firstexampleregion}] 
	The proof follows by evaluating the strong secrecy-distortion region $\mathcal{R}_{\textnormal{POF}}$ defined in Theorem~\ref{theo:SimplifiedforPOF}. Proofs for (\ref{eq:distortion1forexample}) and (\ref{eq:distortion2forexample}) follow by choosing $\Est_j(1,y_j)=y_j$ and $\Est_j(0,y_j)=\mathds{1}\{\Pr[S_j=1]>0.5\}$ for $j=1,2$ that can be obtained as in (\ref{eq:deterministicest}), which are equivalent to the proofs for  \cite[Eqs. (27c) and (27d)]{MariMicheleBCJCAS}. We next have $I(X;Y_1|S_1)=qH_b(p)$, which is equivalent to the proof for \cite[Eq. (27a)]{MariMicheleBCJCAS} with $r=1$. Furthermore, we obtain
	\begin{align}
		&H(Y_1,S_1|Y_2,S_2)-H(S_1|Y_1,Y_2,S_2,X)\nonumber\\
		&\overset{(a)}{=}H(S_1|S_2)+H(Y_1|S_1,Y_2,S_2)-H(S_1|S_2)+I(S_1;Y_1,X|S_2)\nonumber\\
		&\overset{(b)}{=}P_{S_1S_2}(1,0)H(Y_1|S_1=1,S_2=0)+H(X)+H(Y_1|X,S_2)-H(Y_1,X|S_2,S_1)\nonumber
						\end{align}
		\begin{align}
		&\overset{(c)}{=}P_{S_1S_2}(1,0)H(X)+H(X)+P_{X}(1)P_{S_2}(0)H(Y_1|X=1,S_2=0)\nonumber\\
		&\qquad +P_X(1)P_{S_2}(1)H(Y_1|X=1,S_2=1)-H(X)\nonumber\\
		&\overset{(d)}{=}q(1-\alpha)H_b(p)+ p(1-q\alpha)H_b\Big(\frac{q(1-\alpha)}{(1-q\alpha)}\Big)
	\end{align}
	where $(a)$ follows since $S_1-S_2-Y_2$ and $S_1-(Y_1,S_2,X)-Y_2$ form Markov chains for the considered JCAS channel, $(b)$ follows since if $S_1=0$, then $Y_1=0$; if $(S_1,S_2)=(1,1)$, then $Y_1=Y_2=X$; and if $S_2=0$, then $Y_2=0$, and because $X$ is independent of $S_2$, $(c)$ follows since $Y_1=X$ if $S_1=1$, because $X$ is independent of $(S_1,S_2)$, since $Y_1=0$ if $X=0$, and because $(S_1,X)$ determine $Y_1$, and $(d)$ follows since $S_1=1$ if $S_2=1$ due to (\ref{eq:PS1S2exampledistribution}) and because $(S_1,X)$ determine $Y_1$. Therefore, we have
	\begin{align}
		&R\leq \min\Big\{\big(H(Y_1,S_1|Y_2,S_2) - H(S_1|Y_1,Y_2,S_2,X)\big), \quad I(X;Y_1|S_1)\Big\}\nonumber\\
		&= \min\Bigg\{\Bigg(q(1-\alpha)H_b(p)+ p(1-q\alpha)H_b\Big(\frac{q(1-\alpha)}{(1-q\alpha)}\Big)\Bigg), \quad qH_b(p)\Bigg\}.
	\end{align}
\end{IEEEproof}

The securely-transmitted message rate for JCAS scenarios under full secrecy is upper bounded both by $\big(H(Y_1,S_1|Y_2,S_2) -H(S_1|Y_1,Y_2,S_2,X)\big)$ and $I(X;Y_1|S_1)$, the latter of which is the upper bound for the rate when there is no secrecy constraint \cite[Corollary~4]{MariMicheleBCJCAS}. Thus, secrecy might incur a rate penalty for this example. Nevertheless, JCAS methods achieve significantly better performance than separation-based secure communication and state-sensing methods. One can illustrate this by showing that time sharing between the operation point with the maximum secrecy rate and the point with the minimum distortions results in a region that is strictly smaller than the one identified in Lemma~\ref{lem:firstexampleregion}. These analyses are analogous to the comparisons between joint and separation-based secrecy and reliability methods for the secret key agreement problem, as discussed in \cite{bizimWZ,Blochpaper,bizimPeterISITA}.

\section{Proofs for Propositions~\ref{prop:InnerforPSPOF} and \ref{prop:OuterforPSPOF} }\label{sec:proofinnerouterPSPOF}

\subsection{Inner Bound}\label{subsec:innerboundProposition1proof}
\begin{IEEEproof}[Proof Sketch]
	We use the OSRB method \cite{OSRBAmin,RenesRenner} for the achievability proofs, applying the steps in~\cite[Section~1.6]{BlochLectureNotes2018}; see also \cite{YenerOSRB}.
	
	We first define an operationally dual source coding problem to the original JCAS problem, as defined in \cite{OSRBAmin}, along with a coding scheme called \emph{Protocol~A}, for which reliability and secrecy analysis is conducted. We next define a randomized coding scheme, called \emph{Protocol~B}, for the original JCAS problem and show that the joint probability distributions induced by Protocols~A and B are almost equal, which allows to invert the source code proposed for Protocol~A to construct a channel code for Protocol~B. The achievability proof follows by derandomizing the protocols.
	
	\textbf{Protocol~A} (dual source coding problem): We consider a secret key agreement model, in which a source encoder observes $X^n\in\mathcal{X}^n$ and independently and uniformly-randomly assigns three random bin indices $M\in\mathcal{M}=\mathcal{M}_1\times\mathcal{M}_2$ and $F\in\mathcal{F}$ to it. In the dual source model, $M=(M_1,M_2)$ represents a secret key that should be reliably reconstructed at a source decoder that observes $(Y_1^n, S_1^n)\in\mathcal{Y}_1^n\times \mathcal{S}_1^n$ and $F$ to satisfy (\ref{eq:reliability_cons}), whereas the eavesdropper observes $(Y_2^n, S_2^n)\in\mathcal{Y}_2^n\times \mathcal{S}_2^n$ and $F$, which determines the conditions to satisfy the strong secrecy constraint (\ref{eq:secrecyleakage_cons}). Furthermore, the state sequence estimation at the source encoder by using perfect output feedback should satisfy the distortion constraints (\ref{eq:distortion_consts}).
	
	While the strictly causal feedback that depends on the i.i.d. state sequence does not provide opportunities to improve reliability, feedback offers significant opportunities to improve secrecy. We apply a block Markov coding scheme that consists of $b\geq 2$ transmission blocks, each with $n$ channel uses, to transmit $(b-1)$ independent messages $M(k)=(M_{1}(k),M_{2}(k))$. In every block, secret keys are distilled from the feedback and used to protect messages in the subsequent block. In the following, all $n$-letter random variables are  i.i.d. according to (\ref{eq:jointprobiid}) for all $k=[1:b]$,  obtained by fixing $P_{U|V}$, $P_{V|X}$, and $P_X$ so that there exist associated per-letter estimators $\Est_j(x,y_1,y_2)=\widehat{S_j}$ for $j=1,2$ that satisfy 
	\begin{align}
		&\mathbb{E}[d_j(S_j^n,\Est_j^n(X^n,Y^n_1,Y_2^n))]\leq D_j+\epsilon_n^{\prime}\label{eq:assumedperletterestimatorach}
	\end{align}
	where $\epsilon_n^{\prime}>0$ such that $\epsilon_n^{\prime}\rightarrow 0$ when $n\rightarrow\infty$. The block $k$ under consideration is indicated by adding the argument $(k)$ to the variables, e.g., $M(k)$ refers to the message in block $k$, etc.
	
	For all blocks $k=[1:b]$ we construct codes as follows. To each $u^n(k)$ independently and uniformly assign two random bin indices $(F_{\text{u}}(k),W_{\text{u}}(k))$ such that 	$F_{\text{u}}(k)\in[1:2^{n\widetilde{R}_{\text{u}}}]$ and $W_{\text{u}}(k)\in[1:2^{nR_{\text{u}}}]$ for all $k=[1:b]$. Furthermore, to each $v^n(k)$ independently and uniformly assign three random indices $(F_{\text{v}}(k),W_{\text{v}}(k),L_{\text{v}}(k))$ such that $F_{\text{v}}(k)\in[1:2^{n\widetilde{R}_{\text{v}}}]$, $W_{\text{v}}(k)\in[1:2^{nR_{\text{v}}}]$, and $L_{\text{v}}(k)\in[1:2^{n\overline{R}_{\text{v}}}]$ for all $k=[1:b]$. Finally, to each $y_1^{n}(k-1)$, independently and uniformly assign a random index $L_{\text{y}_1}(k\!-\!1)\in[1:2^{n\overline{R}_{\text{y}_1}}]=[1:2^{n\overline{R}_{\text{v}}}]$. Conceptually, the indices $F(k)=(F_{\text{u}}(k), F_{\text{v}}(k))$ represent the public choice of an independent encoder-decoder pair in block $k\in[1:b]$, while the indices $W(k)=(W_{\text{u}}(k), W_{\text{v}}(k),L_v(k))$ represent the messages that should be reliably reconstructed at the decoder. Only $W_{\text{v}}(k)$ should be directly kept secret from the eavesdropper. Moreover, $L_{\text{v}}(k)$ represents a non-secure additional message that should be reliably reconstructed at the decoder and can be kept secret by applying a one-time pad as used in the chosen-secret model \cite{bizimGlobalSIP,IgnaTrans,benimdissertation}. The role of the index $L_{\text{y}_1}(k-1)$, which is known at all legitimate parties thanks to the perfect output feedback, is to provide the required key for the one-time pad in block $k$. Secure reconstruction of $L_{\text{v}}(k)$ follows by summing it in modulo-$2^{n\overline{R}_{\text{v}}}$ with $L_{\text{y}_1}(k\!-\!1)$. Thus, rather than reconstructing $L_{\text{v}}(k)$ directly, the decoder reconstructs the modulo-sum $(L_{\text{v}}(k)\!+\!L_{\text{y}_1}(k\!-\!1))$ by estimating $V^ n(k)$ since it can then use its observation $Y_1^n(k\!-\!1)$ from the previous transmission block to obtain $L_{\text{v}}(k)$ by applying modulo-$2^{n\overline{R}_{\text{v}}}$ subtraction. If $L_{\text{y}_1}(k\!-\!1)$ is uniformly distributed and independent of all random variables in the source model except $Y_1^n(k\!-\!1)$, then the modulo-sum is also uniformly distributed and independent of $L_{\text{v}}(k)$, which allows to keep $L_{\text{v}}(k)$ secret from the eavesdropper. Furthermore, we set for all $k=[2:b]$ that
	\begin{align}
		&M_1(k) = W_{\text{u}}(k), \qquad\quad M_2(k)= (W_{\text{v}}(k), L_{\text{v}}(k))\label{eq:assignmentofM1M2ach}.
	\end{align}
	
	We next impose conditions on the bin sizes to satisfy all constraints given in Definition~\ref{def:systemmodel}.
	
	Using a Slepian-Wolf \cite{SW} decoder, from $(Y_1^n(k),S_1^n(k),F_{\text{u}}(k))$ one can reliably reconstruct $U^n(k)$ for all $k=[1:b]$ such that the expected value of the error probability taken over the random bin assignments vanishes when $n\rightarrow\infty$, if we have \cite[Lemma 1]{OSRBAmin}
	\begin{align}
		\widetilde{R}_{\text{u}} > H(U|Y_1,S_1).\label{eq:reconstrU}
	\end{align}	
	Similarly, one can reliably reconstruct $V^n(k)$ from $(Y_1^n(k),S_1^n(k),F_{\text{v}}(k), U^n(k))$ for all $k=[1:~b]$ if we have
	\begin{align}
		\widetilde{R}_{\text{v}} > H(V|Y_1,S_1,U).\label{eq:reconstrV}
	\end{align}
	Thus, (\ref{eq:reliability_cons}) is satisfied if (\ref{eq:reconstrU}) and (\ref{eq:reconstrV}) are satisfied and \emph{backward decoding} is applied. Backward decoding is a method proposed in \cite{WillemsBackwardDecoding} to decode the blocks in the backward order as $k=b,b-1,\ldots,2$, such that reliable reconstruction of $L_{\text{v}}(k)$ is possible by using $Y_1^n(k\!-\!1)$. 
	
	The public index $F_{\text{u}}(k)$ and secret key $W_{\text{u}}(k)$ are almost independent and uniformly distributed for all $k=[1:b]$ if we have \cite[Theorem 1]{OSRBAmin}
	\begin{align}
		R_{\text{u}}+\widetilde{R}_{\text{u}}<H(U)\label{eq:independenceofFuWu}
	\end{align} 
	since the expected value, taken over the random bin assignments, of the variational distance between the joint probability distributions $\text{Unif}[1\!\!:\!2^{nR_{\text{u}}}]\cdot \text{Unif}[1\!\!:\!2^{n\widetilde{R}_{\text{u}}}]$ and $P_{W_{\text{u}}F_{\text{u}}}$ then vanishes when $n\rightarrow\infty$. Furthermore, the public index $F_{\text{v}}(k)$ and secret key $W_{\text{v}}(k)$ are almost independent of $(Y_2^n(k), S_2^n(k), U^n(k))$ and uniformly distributed for all $k=[1:b]$ if we have 
	\begin{align}
		R_{\text{v}}+\widetilde{R}_{\text{v}}<H(V|Y_2,S_2,U).\label{eq:independenceofFvWv}
	\end{align} 
	Similarly, the random bin index $L_{\text{y}_1}(k-1)$ is almost independent of $\big(Y_2^n(k-1), S_2^n(k-1),$ $ V^n(k-1), U^n(k-1)\big)$ and uniformly distributed for all $k=[2:b]$ if we have 
	\begin{align}
		\overline{R}_{\text{y}_1}&= \overline{R}_{\text{v}} < H(Y_1|Y_2,S_2,V,U)\overset{(a)}{=}H(Y_1|Y_2,S_2,V)\label{eq:independenceofYindexnewwwwwwww}
	\end{align} 
	where $(a)$ follows because $U-V-(Y_1,Y_2,S_2)$ form a Markov chain. Thus, (\ref{eq:secrecyleakage_cons}) is satisfied by applying the one-time padding step mentioned above if (\ref{eq:independenceofFvWv}) and (\ref{eq:independenceofYindexnewwwwwwww}) are satisfied. Consider next the joint condition that $(F_{\text{u}}(k),W_{\text{u}}(k),F_{\text{v}}(k),W_{\text{v}}(k),L_{\text{v}}(k))$ are almost mutually independent and uniformly distributed for all $k=[1:b]$ if we have
	\begin{align}
		R_{\text{u}}+\widetilde{R}_{\text{u}}+R_{\text{v}}+\widetilde{R}_{\text{v}} + \overline{R}_{\text{v}}< H(U,V).\label{eq:sumindependence}
	\end{align}
	
	Applying the Fourier-Motzkin elimination to (\ref{eq:reconstrU})-(\ref{eq:sumindependence}), for any $\epsilon>0$ we can simultaneously achieve 
	\begin{align}
		R_1 &= R_{\text{u}} = I(U;Y_1,S_1) -2\epsilon\overset{(a)}{=} I(U;Y_1|S_1)-2\epsilon\label{eq:R1inProtocolA}
	\end{align}
	and 
	\begin{align}
		&R_2 = R_{\text{v}} +\overline{R}_{\text{v}}\nonumber\\
		& \overset{(b)}{=} \min\{[I(V;Y_1|S_1,U)\!-\!I(V;Y_2|S_2,U)]^{+}\!+\!H(Y_1|Y_2,S_2,V),\quad (I(V;Y_1|S_1)\!-\!R_1)\}\!-\!3\epsilon\label{eq:R2inProtocolB}
	\end{align}
	where $(a)$ follows since $U$ and $S_1$ are independent and $(b)$ follows because $(U,V)$ are mutually independent of $(S_1,S_2)$ and if $H(V|Y_2,S_2,U)\leq H(V|Y_1,S_1,U)$, then $W_{\text{v}}$ cannot be securely reconstructed, i.e., we then impose $R_{\text{v}}=0$.
	
	We next consider the distortion constraints  (\ref{eq:distortion_consts}) on channel-state estimations. Since we assume per-letter estimators given in (\ref{eq:assumedperletterestimatorach}),  (\ref{eq:reliability_cons}) is satisfied by imposing the conditions above on the bin sizes, and all $(u^n(k),v^n(k),x^n(k),y_1^n(k),y_2^n(k),s_1^n(k),s_2^n(k))$ tuples are in the jointly typical set with high probability, by applying the law of total expectation to bounded distortion metrics and from the typical average lemma \cite[pp.~26]{Elgamalbook}, distortion constraints (\ref{eq:distortion_consts}) are satisfied; see also \cite{CorrelatedPaperLong}. Furthermore, without loss of generality one can use the deterministic per-letter estimators in (\ref{eq:deterministicest}) and the proof follows from the proof of \cite[Lemma~1]{MariMichelleGJournalEarlyAccess} by replacing $(S,Z,\hat{S},d)$ with $(S_j,(Y_1,Y_2),\widehat{S}_j,d_j)$, respectively, since $\widehat{S_j}(k)-(X(k),Y_1(k),Y_2(k))-S_j(k)$ form a Markov chain for all $j=1,2$ and $k=[1:b]$.
	
	\textbf{Protocol~B} (random channel coding for the original problem): We consider the original JCAS problem and assist the problem with the public index $F(k)$ for all $k=[1:b]$ such that (\ref{eq:rates_cons})-(\ref{eq:distortion_consts}) are satisfied also for Protocol~B by choosing $R_1$ and $R_2$ as in (\ref{eq:R1inProtocolA}) and (\ref{eq:R2inProtocolB}), respectively. The proof of this result follows by proving that the joint probability distribution obtained in Protocol~A is almost preserved in Protocol~B, i.e., we prove for $n\rightarrow\infty$ that
	\begin{inparaenum}
		\item  the limit of the expectation, defined over the random binning operations, of the variational distance between the joint probability distributions obtained in Protocol~B and required for the reliability constraint is $0$;
		\item the limit of the random probability, defined over the random binning operations, that Kullback-Leibler divergence between the joint probability distributions obtained in Protocol~B and required for the secrecy constraint is greater than $0$ is $0$.
	\end{inparaenum}		
	Since the proof steps are standard and mainly repeat the steps in \cite{OSRBAmin}, we omit them; see \cite[Section~IV]{YenerOSRB} for an extensive proof for a wiretap channel.
	
	Now suppose the public indices $F(k)$ are generated uniformly at random for all $k=[1:b]$ independently. The encoder generates $(U^n(k),V^n(k))$ according to $P_{U^n(k)V^n(k)|X^n(k)F_{\text{u}}(k)F_{\text{v}}(k)}$ obtained from the binning scheme above to compute the bins $W_{\text{u}}(k)$ from $U^n(k)$ and $(W_{\text{v}}(k),L_{\text{v}}(k))$ from $V^n(k)$, respectively, for all $k=[1:b]$. This procedure induces a joint probability distribution that is almost equal to $P_{UVXY_1Y_2S_1S_2}$ fixed above \cite[Section 1.6]{BlochLectureNotes2018}. We remark that the reliability and secrecy metrics considered above are expectations over all possible realizations $F=f$. Thus, applying the selection lemma \cite[Lemma 2.2]{Blochbook}, these results prove Proposition~\ref{prop:InnerforPSPOF} by choosing an $\epsilon>0$ such that $\epsilon\!\rightarrow\! 0$ when $n\!\rightarrow\!\infty$ and imposing $b\!\rightarrow\!\infty$.
\end{IEEEproof}

\subsection{Outer Bound}\label{subsec:outerboundProposition2proof}
\begin{IEEEproof}[Proof Sketch]
	Assume that for some $\delta_n\!>\!0$ and $n\geq 1$, there exist an encoder, decoder, and estimators such that (\ref{eq:rates_cons})-(\ref{eq:distortion_consts}) are satisfied for some tuple $(R_1,R_2,D_1,D_2)$. Using Fano's inequality and (\ref{eq:reliability_cons}), we have
	\begin{align}
		H(M|Y_1^n,S_1^n)\!\overset{(a)}{\leq}\!H(M|\widehat{M})\!\leq\!n\epsilon_n \label{eq:fanoapp} 
	\end{align}
	where $(a)$ allows randomized decoding and $\epsilon_n\!=\!\delta_n (R_1\!+\!R_2)\!+\!H_b(\delta_n)/n$ such that $\epsilon_n\!\rightarrow\!0$ if $\delta_n\!\rightarrow\!0$. 
	
	Let $V_{i}\triangleq (M_1,M_2,Y^{i-1}_{1},S_1^{i-1},Y^{i-1}_{2},S_2^{i-1})$ such that $V_i-X_i-(Y_{1,i},Y_{2,i},S_{1,i},S_{2,i})$ form a Markov chain for all $i\in[1:n]$ by definition of the channel statistics.
	
	\textbf{Bound on $\mathbf{R_1}$}: We have
	\begin{align}
		& nR_1 \overset{(a)}{\leq} I(M_1; Y_1^n|S_1^n)+n\epsilon_n\nonumber\\
		&\leq \sum_{i=1}^n \big(H(Y_{1,i}|S_{1,i})-H(Y_{1,i}|M_1,M_2,Y_1^{i-1},S_1^n)+\epsilon_n\big)\nonumber\\
		& \overset{(b)}{\leq}\sum_{i=1}^n \big(H(Y_{1,i}|S_{1,i})-H(Y_{1,i}|M_1,M_2,Y_{1}^{i-1},S_{1}^{i},Y_2^{i-1},S_2^{i-1})+\epsilon_n\big)\nonumber\\
		&\overset{(c)}{=}\sum_{i=1}^n \big(I(V_i;Y_{1,i}|S_{1,i}) +\epsilon_n)\label{eq:boundonR1converse}
	\end{align}
	where $(a)$ follows by (\ref{eq:fanoapp}) and because $M_1$ and $S_1^n$ are independent, $(b)$ follows since
	\begin{align}
		S_{1,i+1}^{n}-(M_1,M_2,Y_1^{i-1},S_{1}^{i})-Y_{1,i}\label{eq:Markovsi+1}
	\end{align}
	form a Markov chain, and $(c)$ follows from the definition of $V_i$.
	
	\textbf{Bound on $\mathbf{(R_1+R_2)}$}: Similar to (\ref{eq:boundonR1converse}), we obtain
	\begin{align}
		&n(R_1+R_2)\overset{(a)}{\leq} I(M_1,M_2;Y_1^n|S_1^n) +n\epsilon_n\nonumber\\
		&\overset{(b)}{\leq}\sum_{i=1}^n \big(H(Y_{1,i}|S_{1,i})-H(Y_{1,i}|M_1,M_2,Y_{1}^{i-1},S_{1}^{i},Y^{i-1}_{2},S_2^{i-1})+\epsilon_n\big)\nonumber\\
		&\overset{(c)}{=}\sum_{i=1}^n \big(I(V_i;Y_{1,i}|S_{1,i}) +\epsilon_n)
	\end{align}
	where $(a)$ follows because $(M_1,M_2,S_1^n)$ are mutually independent and by (\ref{eq:fanoapp}),  $(b)$ follows since (\ref{eq:Markovsi+1}) form a Markov chain, and $(c)$ follows from the definition of $V_i$.
	
	\textbf{Bound on $\mathbf{R_2}$}: We obtain
	\begin{align}
		&nR_2 \overset{(a)}{\leq}I(M_2;Y_1^n,Y_2^n,S_1^n,S_2^n)+n\epsilon_n\nonumber\\
		&\leq H(Y_{1}^n,S_{1}^n|Y_{2}^n,S_{2}^n)+H(Y_2^n,S_2^n) -H(Y_2^n,S_2^n|M_2)-H(Y_1^n,S_1^n|Y_2^n,S_2^n,M_1,M_2)+n\epsilon_n\nonumber
			\end{align}
		\begin{align}
		&\leq H(Y_{1}^n,S_{1}^n|Y_{2}^n,S_{2}^n)+I(Y_2^n,S_2^n;M_2) -\sum_{i=1}^nH(S_{1,i}|Y_1^n,Y_2^n,S_2^n,M_1,M_2,S_1^{i-1})+n\epsilon_n\nonumber\\
		&\overset{(b)}{\leq}H(Y_{1}^n,S_{1}^n|Y_{2}^n,S_{2}^n)+\delta_n -\sum_{i=1}^nH(S_{1,i}|Y_1^i,Y_2^i,S_2^i,M_1,M_2,S_1^{i-1})+n\epsilon_n\nonumber\\
		&\overset{(c)}{=}H(Y_{1}^n,S_{1}^n|Y_{2}^n,S_{2}^n)+\delta_n -\sum_{i=1}^nH(S_{1,i}|Y_{1,i},Y_{2,i},S_{2,i},V_i)+n\epsilon_n\nonumber\\
		&\leq\sum_{i=1}^n \big(H(Y_{1,i},S_{1,i}|Y_{2,i},S_{2,i})- H(S_{1,i}|Y_{1,i},Y_{2,i},S_{2,i},V_i))\big)+n\epsilon_n+\delta_n
	\end{align}
	where $(a)$ follows by (\ref{eq:fanoapp}), $(b)$ follows by (\ref{eq:secrecyleakage_cons}) and from Remark~\ref{rem:secrecyconstraintwithoutcond}, and because 
	\begin{align}
		(Y_{1,i+1}^{n},Y_{2,i+1}^{n},S_{2,i+1}^{n})- (Y_{1}^i,Y_{2}^i,S_{2}^i,M_1,M_2,S_1^{i-1})-S_{1,i}
	\end{align}
	form a Markov chain, and $(c)$ follows from the definition of $V_i$.
	
	\textbf{Distortion Bounds}: We have for $j=1,2$ 
	\begin{align}
		(D_j\!+\!\delta_n)\overset{(a)}{\geq} \mathbb{E}\big[d_j(S_j^n,\widehat{S_j^n})\big] = \frac{1}{n}\sum_{i=1}^n\mathbb{E}\big[d_j(S_{j,i},\widehat{S_{j,i}})\big] \label{eq:outerbounddistortion}
	\end{align}
	where $(a)$ follows by (\ref{eq:distortion_consts}), which can be achieved by using the deterministic estimators in (\ref{eq:deterministicest}).
	
	Introduce a uniformly distributed time-sharing random variable $\displaystyle Q\!\sim\! \text{Unif}[1\!:\!n]$ that is independent of other random variables, and define $Y_1\!=\!Y_{1,Q}$, $\displaystyle S_1\!=\!S_{1,Q}$, $\displaystyle Y_2\!=\!Y_{2,Q}$, $\displaystyle S_2\!=\!S_{2,Q}$, $X\!=\!X_Q$, and $V\!=\!(V_{Q},\!Q)$, so $V-X-(Y_1,Y_2,S_1,S_2)$ form a Markov chain. The proof of the outer bound follows by letting $\delta_n\rightarrow0$.
	
	\textbf{Cardinality Bounds}: We use the support lemma \cite[Lemma 15.4]{CsiszarKornerbook2011} to prove the cardinality bound, which is a standard procedure, so we omit the proof.
\end{IEEEproof}

\bibliographystyle{IEEEtran}
\bibliography{references}
\end{document}